\documentstyle[aps,epsf,pre]{revtex}
\begin{document}
\title{Non-normal parameter blowout bifurcation: an example in a 
truncated mean field dynamo model}
\author{Eurico Covas${}^{1}$\thanks{E-mail:E.O.Covas@qmw.ac.uk},
Peter Ashwin${}^{2}$\thanks{E-mail:P.Ashwin@mcs.surrey.ac.uk}
and Reza Tavakol${}^{1}$\thanks{E-mail:reza@maths.qmw.ac.uk}}
\address{1. Astronomy Unit                           \\
School of Mathematical Sciences                   \\
Queen Mary \& Westfield College                   \\
Mile End Road                                     \\
London E1 4NS, UK                                 \\
and                                               \\
2. Department of Mathematical and Computing Sciences \\
University of Surrey                              \\
Guildford GU2 5XH, UK                             }
\date{\today}
\maketitle
\begin{abstract}
We examine global dynamics and bifurcations occurring in a truncated
model of a stellar mean field dynamo.  This model has symmetry-forced invariant
subspaces for the dynamics and we find examples of transient type I
intermittency and blowout bifurcations to transient on-off
intermittency, involving laminar phases in the invariant submanifold.
In particular, our model provides examples of
blowout bifurcations that occur on varying a non-normal parameter; 
that is, the parameter varies the dynamics within the invariant
subspace at the same time as the dynamics normal to it. As a consequence
of this we find that the Lyapunov exponents do not vary smoothly and
the blowout bifurcation occurs over a range of parameter values rather
than a point in the parameter space.

\end{abstract}
\newcommand{\COM}[1]{COMMENT: {\bf {#1}}}
\section{Introduction}
There has recently been a great deal of work on deterministic dynamical
systems possessing invariant submanifolds, motivated by symmetric
systems, and in particular some coupled systems. Such systems
have been shown to be
capable of producing a range of interesting and robust dynamical modes of
behaviour, such as {\it riddled basins} \cite{alexander-etal} and {\it
on-off intermittency} \cite{platt}, shown to be related by the {\it
blowout bifurcation} \cite{ott-sommerer}, as well as {\it transient
on-off intermittency} \cite{xie-hu}.  We denote the manifold on which
such a system is defined by $M$ and the corresponding invariant
submanifold by $M_I$. An attractor is said to have a riddled basin
if every open set intersecting the basin also intersects the basin of
another attractor in a set of positive measure; such basins can arise
robustly for Milnor attractors \cite{Milnor} contained in $M_I$.

On-off intermittency to a state in $M_I$ is characterised by dynamics that comes
arbitrarily close to the state in $M_I$ but that also has intermittent large
deviations from $M_I$.  Transient on-off intermittency is a
transient dynamics exhibiting characteristics of on--off intermittent behaviour.
Namely, it has an average distribution of laminar phases that satisfies a
power law with exponent $-3/2$ \cite{xie-hu}. All these phenomena arise
as different aspects of blowout bifurcations, where a maximum normal
Lyapunov exponent of an attractor in $M_I$ passes through zero and thus
causes a loss of stability. To date, most mathematical understanding of
such systems is limited to cases where there are a number of
simplifying assumptions:

\begin{itemize}
\item[(H1)]
The control parameters are assumed to be {\it normal}
\cite{ashwin-etal}, in the sense that the dynamics of the invariant
submanifold is independent of the bifurcation parameter. Such
parameters preserve the dynamics on the invariant submanifold, but allow
it to vary in the rest of the phase space;
\item[(H2)]
The attractor which becomes transversely unstable in the $M_I$ and hence
causes the blowout bifurcation is usually assumed to be chaotic.
\end{itemize}
(Note however that \cite{lai-T2} find blowout type bifurcations from quasiperiodic
dynamics on $T^2$.)

As many physical systems of interest are unlikely to possess normal
parameters (a notable exception to this being some coupled systems), we
expect that (H1) is not usually applicable.  This is
particularly expected to be true in truncations of partial differential
equations that arise in fluid and dynamo equations (as well as
astrophysical models in general). Thus, by studying the behaviour of
a specific example where (H1) does not hold, we hope we throw some light on
the dynamics of general systems of this type. 

To this end we examine a system where (H1) does not hold,
i.e.\ where the control parameter varies the dynamics within the invariant
subspace as well as that normal to it. We see that this has the effect of
`spreading' the blowout bifurcation out over an interval of parameter
values due to the existence of periodic windows where (H2) does not hold;
however, we conjecture that there is a positive measure
subset of parameters on which the blowout resembles the case for
normal parameters.

The model we describe in Section~II arises as the truncation of a
stellar axisymmetric mean field dynamo model where 
the natural control parameters are not normal. There is also more than
one invariant manifold forced by the spatial symmetries of the system,
although this does not appear to affect the behaviour in the cases
examined, in the sense that it is only one of these manifolds, namely
the antisymmetric one, which seems to dominate the attracting dynamics. 

In Section~III we discuss numerical results from simulations of the model; we
discuss the basic bifurcational behaviour in the full system and the antisymmetric
subspace before discussing examples of type I intermittency, transient on--off
intermittency and non-normal parameter blowout bifurcation in the model.
In the final Section~IV the implications of the results are discussed for more general
systems of this type as well as for the dynamo problem.

\section{Model}

\subsection{Galerkin model for the mean field dynamo}

The dynamo model considered here is the so called $\alpha \Omega$ mean
field dynamo model, with a dynamic $\alpha$--effect, given by Schmalz \&
Stix \cite{schmalzetal91} (see also Covas et al.\ \cite{covasetal96} for
details). We assume a spherical axisymmetrical configuration with one
spatial dimension $x$ (corresponding to the latitude and measured in
terms of the stellar radius $R$) for which the magnetic field takes the
form
$$
\vec{B}=\left(0,B_{\phi},\frac{1}{R}\frac{\partial A_{\phi}}{\partial x}\right),
$$
where $A$ is the $\phi$--component (latitudinal) of the magnetic
vector potential and $B$ is the $\phi$--component of $\vec{B}$.
The model is given by
the mean field induction equation
\begin{equation}\label{induction}
\frac{\partial\vec{B}}{\partial t}=\nabla\,\times\,(\vec{v}\,\times\,\vec{B}+
\alpha\vec{B}-\eta_t\nabla\,\times\,\vec{B}),
\end{equation}
where $\vec{B}$ is the mean magnetic field, $\vec{v}$ is the mean
velocity, $\eta_t$ is the turbulent magnetic diffusitivity
and $\alpha$ is the coefficient of the $\alpha$--effect \cite{krause80}.
In addition the $\alpha$--effect, which is
important in maintaining the dynamo action by relating the mean
electrical current arising in helical turbulence to the mean magnetic
field,
is assumed to be dynamic and expressible in the form
$\alpha=\alpha_0\cos x-\alpha_M(t)$, where $\alpha_0$ is a constant and
$\alpha_M$ is its dynamic part satisfying the equation
\begin{equation}\label{dynamicalpha}
\frac{\partial \alpha_M}{\partial t}= \nu_t
\frac{\partial^2 \alpha_M}{\partial x^2} + Q\,\vec{J}\cdot\vec{B},
\end{equation}
where $Q$ is a physical constant, $\vec{J}$ is the electrical current
and $\nu_t$ is the turbulent diffusivity.
These assumptions allow Eq.\ (\ref{induction}) to be split into the
following two equations:
\begin{eqnarray}
\label{p1}
\frac{\partial A_{\phi}}{\partial t}&=&\frac{\eta_t}{R^2}
\frac{\partial^2A_{\phi}}{\partial x^2}+\alpha B_{\phi},\\
\label{p2}
\frac{\partial B_{\phi}}{\partial t}&=&
\frac{\eta_t}{R^2}\frac{\partial^2 B_{\phi}}
{\partial x^2}+\frac{\omega_0}{R}\frac{\partial A_{\phi}}{\partial x}.
\end{eqnarray}
Expressing these equations in a
non-dimensional form, relabelling the new variables to
$$
(A_\phi,~B_\phi,~ \alpha_M) \Longrightarrow (A,~B,~C),
$$
and using a spectral expansion of the form
\begin{eqnarray*}
A=\sum_{n=1}^{N}A_n(t)\sin nx,\\
B=\sum_{n=1}^{N}B_n(t)\sin nx,\\
C=\sum_{n=1}^{N}C_n(t)\sin nx,
\end{eqnarray*}
where $N$ determines the truncation order, reduces equations
(\ref{dynamicalpha})--(\ref{p2}) to a set of ordinary
differential equations, the
dimension of which depends on the truncation order $N$.
We consider the full system given in terms of the
variables $A_n$, $B_n$, $C_n$, $n=1,\cdots,N$ in the form
\begin{eqnarray}
\label{main3_1}
\frac{\partial A_n}{\partial t}&=&-n^2A_n+\frac{D}{2}(B_{n-
1}+B_{n+1})\\&&+\sum_{m=1}^{N}\sum_{l=1}^{N}F(n,m,l)B_mC_l,\nonumber\\
\label{main3_2}
\frac{\partial B_n}{\partial t}&=&-n^2B_n+\sum_{m=1}^{N}G(n,m)A_m,\\
\label{main3_3}
\frac{\partial C_n}{\partial t}&=&-\nu n^2 C_n
-\sum_{m=1}^{N}\sum_{l=1}^{N}H(n,m,l)A_mB_l,
\end{eqnarray}
where
\begin{eqnarray*}
&F(n,m,l)=&\\&\frac{8nml}{\pi(n+m+l)(n+m-l)(n-m+l)(n-m-l)}&\nonumber,
\end{eqnarray*}
if $n+m+l$ is odd and $F(n,m,l)=0$ otherwise,
\begin{eqnarray*}
&H(n,m,l)=&
\\&\frac{4}{\pi} {\frac {{n} {m} {l}  (- {n}^{2} + 3 {m}
^{2} + {l}^{2} )}{( {n} + {l} + {m} ) ( {n} + {l} - {m} ) ( {
n} - {l} + {m} ) ( {n} - {l} - {m} )}}&\nonumber,
\end{eqnarray*}
if $n+m+l$ is odd and $H(n,m,l)=0$ otherwise and
\begin{equation}
G(n,m)=\frac{4nm}{\pi(n^2-m^2)},
\end{equation}
if $n+m$ is odd and $G(n,m)=0$ otherwise.  In these equations the
control parameters are the so called dynamo number $D$ (which is
proportional to the square of the angular velocity gradient and to the
square of the turnover time of the turbulent convection eddies) and the
diffusivity ratio $\nu=\frac{\nu_t}{\eta_t}$.

\subsection{Invariant subspaces for the model with $N=4$}

Covas et al.\ \cite{covasetal96} confined themselves to the study of
models that are antisymmetric with respect to the equator and found
that the minimum truncation order $N$ for which a similar asymptotic
behaviour existed was $N=4$. In this case, the equations have a twelve 
dimensional phase space and are symmetric under the four-element
Abelian group that comprises the identity $I$, the reversal transformation
$$
A_n(t) \to -A_n(t), \quad B_n(t) \to -B_n(t), \quad C_n(t) \to C_n(t),
$$
the antisymmetric (or dipolar) transformations
$$
A_n(t) \to (-1)^{n+1}A_n(t), \quad B_n(t) \to (-1)^{n}B_n(t), \quad
C_n(t) \to (-1)^{n}C_n(t)
$$
and the symmetric (or quadrupolar) transformations
$$
A_n(t) \to (-1)^{n}A_n(t), \quad B_n(t) \to (-1)^{n+1}B_n(t), \quad
C_n(t) \to (-1)^{n}C_n(t).
$$

The trivial solution, given by $A_n=B_n=C_n=0$,
is the only one which possesses both the dipolar
and quadrupolar symmetries while symmetric solutions come in pairs and
asymmetric solutions come in quadruples.

The antisymmetric part of these equations, which is obtained by putting
$B_1 = C_1= A_2 = B_3 = C_3 = A_4 =0$, was studied in \cite{ecrt97}.
We refer to this dynamically invariant subspace
$$
M_A=\{ (A_1,0,0,0,B_2,C_2,A_3,0,0,0,B_4,C_4)\}
$$
as the antisymmetric subspace. There is also a six dimensional 
symmetric invariant subspace
$$
M_S=\{ (0,B_1,0,A_2,0,C_2,0,B_3,0,A_4,0,C_4)\}
$$
although as we will see, the attractors are typically not contained 
within $M_S$. Throughout the paper we refer to the full system as the
12 -- dimensional system.
\section{Dynamical behaviour}

The system considered here has a two dimensional parameter space $(D,
\nu)$, neither of which is normal for the system restricted to $M_A$, as can be
seen from equations (\ref{main3_1})-(\ref{main3_3}).  We confine $\nu$
to the range $[0,1]$ on physical grounds, as otherwise there will be no
dynamo action.  Previous studies of these models have taken $\nu =0.5$.
Here we shall consider two distinct cases of $\nu$ given by $0.5$ and
$0.47$ in this range and in each case allow $D$ to vary.  To study the
dynamics of this system, we start by looking at the dynamics on the
antisymmetric invariant submanifold $M_A$ and then look
at how this changes as the full system is switched on.  

\subsection{Basic bifurcation behaviour}

To begin with, we consider the case of $\nu=0.5$ and as a first step
make a coarse study of the dynamics confined to $M_A$ as well as the
full (12--dimensional) system by considering the averaged energy ($E
\propto \int_0^{\pi}|\vec{B}|^2 dx$) as a function of the parameter $D$.
The results of these calculations are summarised in Figs.\
(\ref{bifurcation_anti}) and (\ref{bifurcation_mixed}) respectively.
The figures were produced using a fourth order variable step size
Runge-Kutta method to integrate a number of randomly selected initial
conditions forward in time, and so get a selection of the possible
attractors. After a time when transients were deemed to have decayed
(which we took to be 1000 time units) we averaged the energy over a
much longer time series i.e.\ 10000 time units. We have verified the
following results using the continuation package AUTO97 \cite{auto97}.

For small $D$ ($D<98.67$) all attracting dynamics of the
full 12--dimensional system is confined to the 6--dimensional
antisymmetric invariant submanifold $M_A$. The details of bifurcations are
depicted in Figs.\ (\ref{bifurcation_anti}) and
(\ref{bifurcation_mixed}).  As can be seen from these figures, as $D$ is
increased, the fixed point at the origin (the trivial solution for both
systems) bifurcates at $D=12.57$ to two fixed points, which are
symmetric with respect to $A_n\to -A_n$, $B_n\to -B_n$, $C_n\to C_n$. A
subcritical pitchfork-type bifurcation stabilises
the origin from $D=77.25$ up to a supercritical Hopf bifurcation at
$D=98.67$.

For larger values of $D$ the two systems will evolve differently. In 
$M_A$ the supercritical Hopf bifurcation creates a
stable periodic solution while the 12--dimensional system evolves on a
branch of stable periodic solution with quadrupolar symmetry that
becomes unstable through a torus bifurcation at $D=118.76$.  In addition
to this, at $D=118.23$, the origin undergoes Hopf bifurcation creating a branch
of dipolar periodic solutions that are stabilised by a torus bifurcation
at $D=127.78$, as shown in the inset of Fig.\ (\ref{bifurcation_mixed}).
Also shown in the inset is the appearance through a saddle-node
bifurcation of a branch of periodic solutions in $M_S$ at $D=132.72$
that are stable up to $D=135.42$. Here they lose stability at a
torus bifurcation. 
All of these branches of solutions pass through the region from $D=150$ 
to $D=175$ without bifurcation; however they do not provide a complete picture 
of all that happens in this region.

There is a pair of periodic orbits in $M_A$ that
are created at a saddle-node bifurcation at $D=170.25$. These are stabilised
by a pitchfork bifurcation of periodic orbits at $D=171.10$ creating 
a branch of stable periodic orbits with no symmetry continuing down to 
$D=171.003$; these are important for the intermittent dynamics discussed
in the next subsection  (Fig.\ \ref{typeI.bifurcation}).
For larger $D$, the periodic orbit undergoes a sequence of bifurcations 
preserving the dipolar symmetry (i.e.\ creating branches that 
remain within $M_A$) up until $D=177.75$ where a branch of stable periodic
orbits that bifurcate from $M_A$ is created. We conjecture that the saddle
node bifurcation creating this series of periodic orbits is associated
with breakdown of a quasiperiodic flow on a two-torus created at one of
the torus bifurcations but have not been able to check this.

\subsection{Intermittent dynamics}

We investigated two cases; $\nu=0.5$ and $\nu=0.47$. Note that for
physically meaningful results we require $\nu\leq 0.5$.
We have examined the transverse stability of attractors in $M_A$
by calculating the corresponding transverse Lyapunov exponent ($\lambda_T$).
Fig.\ \ref{transverse_nu_0.50} summarises the results of calculations
of the $\lambda_T$ for the periodic orbit which is created at 
$D=170.25$, as shown in the inset in Fig.\ (\ref{bifurcation_anti}).  The
important feature in this behaviour of $\lambda_T$ is the presence
of the two crossings through zero.  The transverse stability of the
other attractors do not change in this range of $D$. 

\subsubsection{The case of $\nu=0.5$}

For the case $\nu=0.5$ we examined the behaviour of the full system over a
parameter interval in the neighbourhood of $D=171$. The loss of stability
of the periodic orbit described in the previous
section, does not induce on-off intermittency as we first suspected.
Instead, the bifurcation at $D=171.10$ is a pitchfork bifurcation that
creates an asymmetric periodic orbit which survives up to $D=171.003$
and is then destroyed by a collision with an unstable orbit in a saddle
node bifurcation, which is shown in Fig.\ \ref{typeI.bifurcation}.

For $D<171.003$, we have a transient type-I intermittency, as can be
seen in Fig.\ \ref{series_nu=0.50}. We also calculated the scalings of
the transient times and average times between the bursts (as shown in
Fig.\ \ref{transients_nu=0.50} and Fig.\ \ref{bursts_nu=0.50}) and found
them to be in good agreement with the known $-1/2$ scaling.  The
behaviour between bursts shown in Fig.\ \ref{series_nu=0.50}, resembles
the 12D periodic orbit discussed above, except that the amplitude of the
symmetric part of the oscillations between the bursts grow slowly
towards the bursts and return, after the bursts, close to the invariant
submanifold. The intermittent behaviour is transient, in the sense that
the orbit returns to the fixed point in the invariant submanifold.  The
spectrum of Lyapunov exponents for these transient forms of
intermittency is in the form $(+,0,-\ldots)$, until the transient dies
out, becoming attracted to a stable fixed point (note: there are two
such fixed points, located symmetrically with respect to the A and B
variables). 

On the negative side of the crossing of the transverse Lyapunov
exponent, shown in Fig.\ \ref{transverse_nu_0.50}, we observe a basin
boundary for the full 12--dimensional system with a dimension close to
that of the phase space.  This is shown in Fig.\ \ref{false_riddled}
which demonstrates which asymptotic attractor on the invariant
submanifold the initial conditions get attracted to.  Both basins seem
to be made up of open sets (supported by the fact that calculations
indicate an integer box counting dimension).  This conclusion is further
supported by the calculation of the exterior dimension \cite{ext} shown
in Fig.\ \ref{exterior}.

It is also interesting to note, from both physical and mathematical
points of view, that even if the full (12--dimensional) system
does have any new attractors, nevertheless it will in general alter
the relative size of the basins of attraction;
most initial points seem to get attracted to only the fixed
points or one of the cycles, as opposed to initial conditions 
starting in the invariant sub-manifold.  

\subsubsection{The case of $\nu=0.47$}

By making $\nu$ slightly smaller than $0.5$ we were 
able to change the order of the bifurcation to chaos in the
invariant submanifold relative to the loss of transverse stability.
To study the behaviour of the system
with $\nu=0.47$, we looked at the parameter region in which the system
with $\nu =0.5$ had a chaotic attractor (as depicted in Fig.\
\ref{transverse_nu_0.50}).  For our calculations, we chose our initial
conditions to lie in the basin of the chaotic attractors for $\nu =0.5$
system.  We then studied the evolution of the system for $\nu=0.47$ by
changing the control parameter and taking the initial conditions at each
step to lie in the basin of the attractor for the previous parameter
value. The results of these calculations for the $\lambda_T$ and the
largest Lyapunov exponents of the full 12--dimensional system are given
in Fig.\ \ref{transverse_nu_0.47} and Fig.\ \ref{amplify}.  As can be
seen, the chaotic behaviour is now interspersed with periodic windows.
Within these windows the periodic solutions co-exist with chaotic
repellers. 

Another crossing of the transverse Lyapunov exponent, from negative to
positive, shown in Fig.\ \ref{transverse_nu_0.50}, occurs at $D=178.71$,
and for $177.10<D<178.71$ there are stable periodic orbits restricted to
the invariant submanifold. This crossing is also related to a
bifurcation of periodic orbits and therefore there is no indication of
on-off intermittency or riddled basins. The periodic orbit on the
invariant submanifold becomes chaotic just after $D=178.76$, not close
enough to the transverse stability bifurcation at $D=177.71$ to induce
on-off intermittency. This suggests that there is likely to be a blowout
at nearby parameters in the two parameter space, as we discuss in the
next section.
To substantiate this, we calculated the scaling of the probability
distribution of the off phases (corresponding to when the distance to
the invariant submanifold is less than $10^{-3}$) as a function of their
length. As can be seen in Fig.\ \ref{laminar}, the scaling agrees with
the power law behaviour proposed in \cite{platt} with an index of
$-3/2$. At this parameter value transient on-off trajectories appear to
be induced by a chaotic invariant set that is a repeller within the
invariant submanifold. 

\subsection{Generic behaviour for non-normal parameters}

There are a number of interesting dynamical phenomena that occur here,
that are related to the fact that the system parameters are 
not normal.\\

(I) The chaotic behaviour in the invariant submanifold appears to be of
the non-uniformly hyperbolic variety, and in particular the chaotic
attractors are not structurally stable; they are destroyed by
arbitrarily small perturbations. Notwithstanding this, we find numerical
evidence (Fig.\ \ref{transverse_nu_0.47} and Fig.\ \ref{amplify}) that
there is a family of chaotic attractors with similar properties defined
on a {\em subset} of parameter space with positive (Lebesgue) measure
but open (even dense) complement. This is what is found, for example,
in the logistic map \cite{jakobson}. In the open dense complement we
expect to see periodic windows and many bifurcations, for example period
doubling cascades, which we have found numerically.  This is consistent with
the conjecture of Barreto et al.\ \cite{barreto} on noting that the
attractors here have only one positive Lyapunov exponent.  In this
parameter region the system may be said to be fragile \cite{rtge}, in
the sense that arbitrarily small changes in the control parameter $D$ 
can force a chaotic attractor to be replaced by a nearly attracting periodic
orbit.

(II) In the light of (I), there is no reason why there should be
a unique parameter value $D_c$ at which blowout occurs. In particular,
the attractor in $M_A$ varies discontinuously, and its tangential and
normal Lyapunov exponents vary
discontinuously with $D$ except within the periodic windows.  This
explains the presence of smooth segments in the curves of Fig.\
\ref{amplify} within regions where the attracting dynamics is periodic.

(III) In this system the passage of $\lambda_T$ through zero is
fairly simple. Firstly, the value of $\lambda_T$ can be bracketed
between upper and lower bounds that also pass through zero. This is
presumably due to the fact that certain periodic orbits in the attractor
will typically maximise and minimise transverse Lyapunov exponents
\cite{hunt-ott}. Secondly, there is evidence that there is a positive
measure Cantor set $\cal{S}$ in parameter space where chaotic
behaviour exists.  On $\cal{S}$ the Lyapunov exponents are continuous in
the sense that there is a continuous function $\lambda'$ of parameter
that is equal to $\lambda_T$ on $\cal{S}$ and passes through zero at
about $D=177.75$ (see Fig.\ \ref{transverse_nu_0.47} and also Fig.\
\ref{average}).  Even for parameter values not in $\cal{S}$ we can get
transient on-off intermittent behaviour (see Fig.\
\ref{series_nu=0.47}). \\

On the basis of our results we conjecture that properties (I)-(II)
are typical behaviour at blowout on varying a non-normal parameter and
(III) is a typical simple scenario of how this can occur.

For the case $\nu=0.47$ we note that the transverse Lyapunov exponent 
(of the chaotic invariant set that attracts within $M_A$) becomes
positive and causes the appearance of transient on-off intermittency.

\section{Discussion}

We have studied global dynamics and bifurcations occurring in a
12--dimensional truncation of a stellar mean field dynamo model which
possesses two six dimensional invariant submanifolds corresponding to
dipolar and quadrupolar symmetries respectively.  An essential feature
of this model is that its control parameters are non-normal, allowing
the dynamics to vary both within the invariant submanifolds as well as
in the directions normal to them. 

Depending upon the region of the parameter space considered, we find a
diverse set of dynamical modes of behaviour, including different forms
of intermittency.  In addition to transient type I intermittency, we
find transient on-off intermittency induced by blowout bifurcations. In
the parameter range where we observe the latter behaviour, the invariant
submanifold possesses a family of chaotic attractors on a subset of
parameter space with positive (Lebesgue) measure but open (even dense)
complement.  On the basis of our numerical calculations these attractors
seem to be structurally unstable, which is consistent with the
conjecture of Barreto et al.\ \cite{barreto}. We also find that as a
consequence of the non-normality of the control parameters the blowout
bifurcation seems to occur over an interval rather than a point in the
parameter space. 

These results can be of potential significance for the dynamical
behaviour of systems with non-normal parameters.  Given the fact that
the model considered here was derived directly from dynamo equations,
the forms of intermittency found here can also be of potential
importance in understanding the mechanism of production of the
so-called grand or Maunder-type minima in solar and stellar activity,
during which the amplitudes of stellar cycles are greatly diminished
\cite{weiss,spiegel}.  We do not however wish to imply that the forms of
intermittency responsible for such stellar behaviour are necessarily
transient.

\acknowledgments

EC is supported by grant BD/5708/95 -- Program PRAXIS
XXI, from JNICT -- Portugal. PA is partially supported by
a Nuffield ``Newly appointed science lecturer'' grant.
RT benefited from PPARC UK Grant No. H09454. This research also benefited
from the EC Human Capital and Mobility (Networks) grant ``Late type stars:
activity, magnetism, turbulence'' No. ERBCHRXCT940483.


\pagebreak

\begin{figure}
\caption{\label{bifurcation_anti}
Diagram showing attractors for a random selection of
initial conditions within $M_A$ for $\nu=0.5$. TFP stands for trivial 
fixed point, FP for non-trivial fixed point, PA for antisymmetric 
periodic orbit and CA for antisymmetric chaotic orbit. Continuation
using DSTOOL shows that the break in the FP attractor uppermost in this
diagram is just a feature of the choice of initial conditions; in fact it
continues to be attracting over the whole range of $D$. The inset 
shows coexisting chaotic and periodic attractors over a range of $D$.}
\end{figure}
\begin{figure}
\caption{\label{bifurcation_mixed}
Diagram showing attractors for randomly chosen initial conditions in
the full phase space for $\nu=0.5$. PS stands for symmetric periodic orbit
(i.e.\ in $M_S$), PM for a periodic orbit neither in $M_A$ nor $M_S$ and
QPA for antisymmetric quasiperiodic orbit. The notation of TFP, FP and PA are
as in Fig.\ {\protect \ref{bifurcation_anti}}. Observe the existence of
intermittent behaviour over a range of $D$.}
\end{figure}
\begin{figure}
\caption{\label{transverse_nu_0.50}
The leading (i.e.\ most positive) transverse Lyapunov exponent and the
two leading Lyapunov exponents for the attractor of a particular initial
condition for the system on $M_A$ at $\nu=0.5$. The attractor is a
periodic orbit from $D=170.25$ up to $D=178.76$, although it undergoes a 
number of period doublings in this range to give a chaotic attractor
for $D>178.76$. The computed orbit is transversely stable in the
range $171.10<D<178.71$. By reducing $\nu$ to $0.47$ we can change order
of the loss of transverse stability and the breakdown to chaos in $M_A$.}
\end{figure}
\begin{figure}
\caption{\label{typeI.bifurcation}
Continuation of a periodic orbit showing
breakdown to type I intermittency at $\nu=0.5$. The abscissa shows a
symmetric component of a branch of PM periodic orbits created at a
pitchfork of a PA periodic orbit. This is destroyed at a saddle-node
bifurcation giving rise to type I intermittency at $D<170.003$.}
\end{figure}
\begin{figure}
\caption{\label{series_nu=0.50}
Time series showing a component transverse to $M_A$ for transient
type I intermittency series for $\nu=0.5$ at $D=170$. Observe the
long but irregular periods of lingering near a small amplitude periodic 
orbit interspersed by large fluctuations. After a long time, the trajectory
is asymptotic to a stable fixed point.}
\end{figure}

\begin{figure}
\caption{\label{transients_nu=0.50}
Scaling of transient time of the transient type I intermittency for $\nu=0.5$ 
against the unfolding parameter $D-D_c$.}
\end{figure}
\begin{figure}
\caption{\label{bursts_nu=0.50}
Scaling of transient burst time of the transient type I intermittency 
for $\nu=0.5$.}
\end{figure}
\begin{figure}
\caption{\label{false_riddled}
Two dimensional slice through phase-space obtained by setting all components
zero except for $A_1$ and $B_1$. The basins of attraction of the fixed
point (black) and the periodic orbit (white) that are coexisting attractors
at $D=171.12$ and $\nu=0.5$.}
\end{figure}
\begin{figure}
\caption{\label{exterior}
Approximation of the exterior dimension $D_x$ of
the basin shown in black in
Fig.\ {\protect \ref{false_riddled}. This is very close to the dimension
of the slice through phase-space indicating that the basin boundary
is highly convoluted, even though it is not riddled.}
}
\end{figure}
\begin{figure}
\caption{\label{transverse_nu_0.47}
Largest transverse Lyapunov exponent ($\lambda_T$) and the two leading
Lyapunov exponents ($\lambda_1$ and $\lambda_2$) for a family of attractors
on the antisymmetric invariant submanifold for $\nu=0.47$. Note the
existence of periodic windows and general trend of $\lambda_T$ through
zero indicate a blowout bifurcation near $D\approx 177.75$. The lack
of smoothness of these curves is indicative of the fact that $D$ is not 
a normal parameter.}
\end{figure}
\begin{figure}
\caption{\label{amplify}
Amplification of the transverse Lyapunov exponent
and the two leading Lyapunov exponents
for the antisymmetric subset of equations for $\nu=0.47$. This shows the
existence of a
``window'' in parameter space where the attractor within the invariant
subspace is periodic and transversely repelling.}
\end{figure}
\begin{figure}
\caption{\label{series_nu=0.47}
Time series showing transient on-off intermittency 
for $\nu=0.47$ and $D=177.70$. The on-off intermittent behaviour is induced by
a chaotic repeller which is present within the periodic windows. After a long 
transient, the trajectory is asymptotic to a stable fixed point within $M_A$ 
(not shown).}
\end{figure}
\begin{figure}
\caption{\label{laminar}
Scaling of the laminar phases over an on-off transient orbit segment for
$\nu=0.47$ and $D=177.70$. The $-3/2$ scaling is evidence of an
on-off intermittent state.}
\end{figure}

\begin{figure}
\caption{\label{average}
Average of the variable $B_2$ measuring average distance from $M_A$ 
over an on-off transient orbit segment that eventually ends at a fixed
point (for $\nu=0.47$). The discontinuous nature of this presumably reflects 
the discontinuous change in $\lambda_T$ illustrated in 
Figure~{\protect \ref{transverse_nu_0.47}}.}
\end{figure}


\centerline{\def\epsfsize#1#2{0.55#1}\epsffile{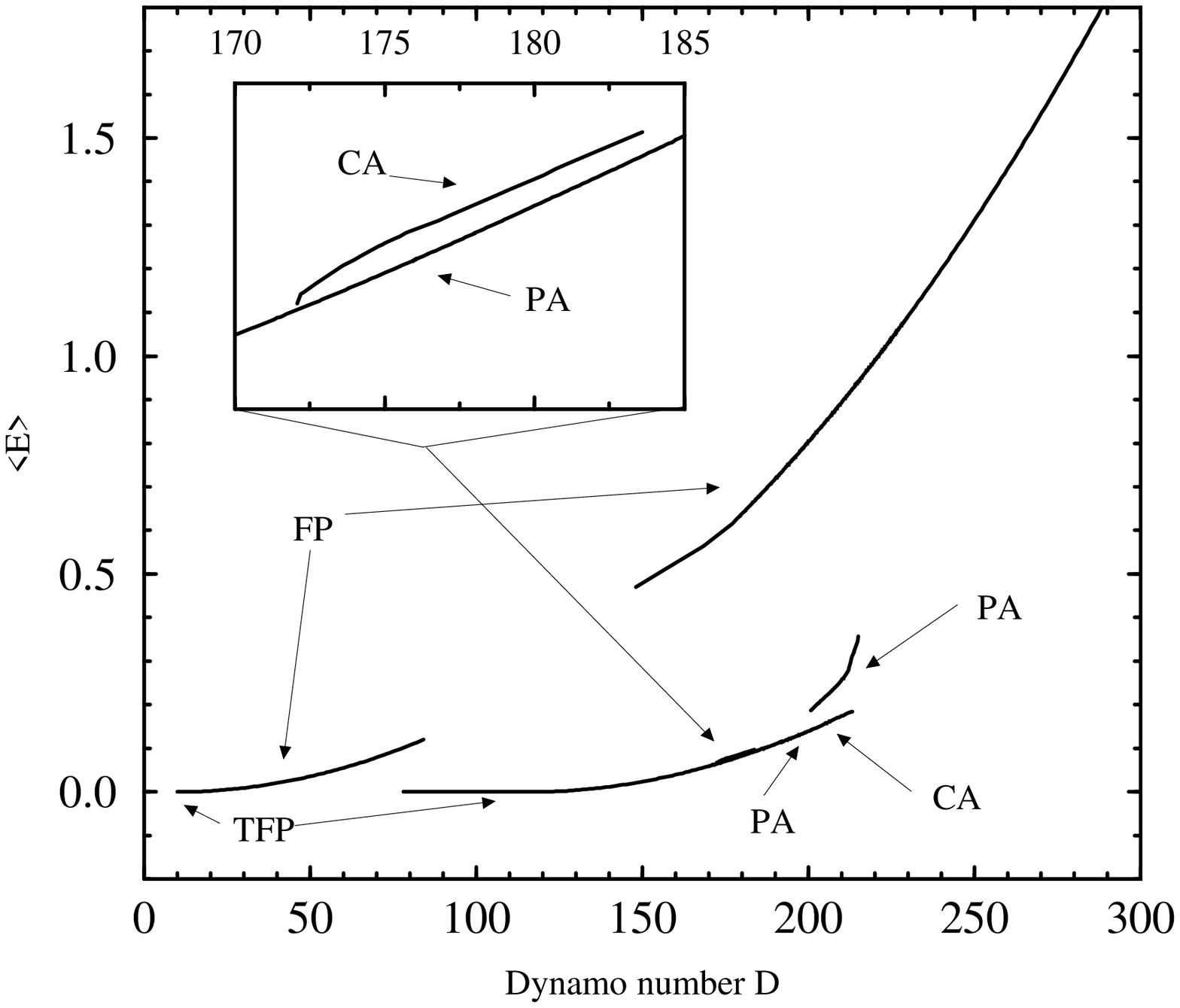}}
Figure 1

\centerline{\def\epsfsize#1#2{0.55#1}\epsffile{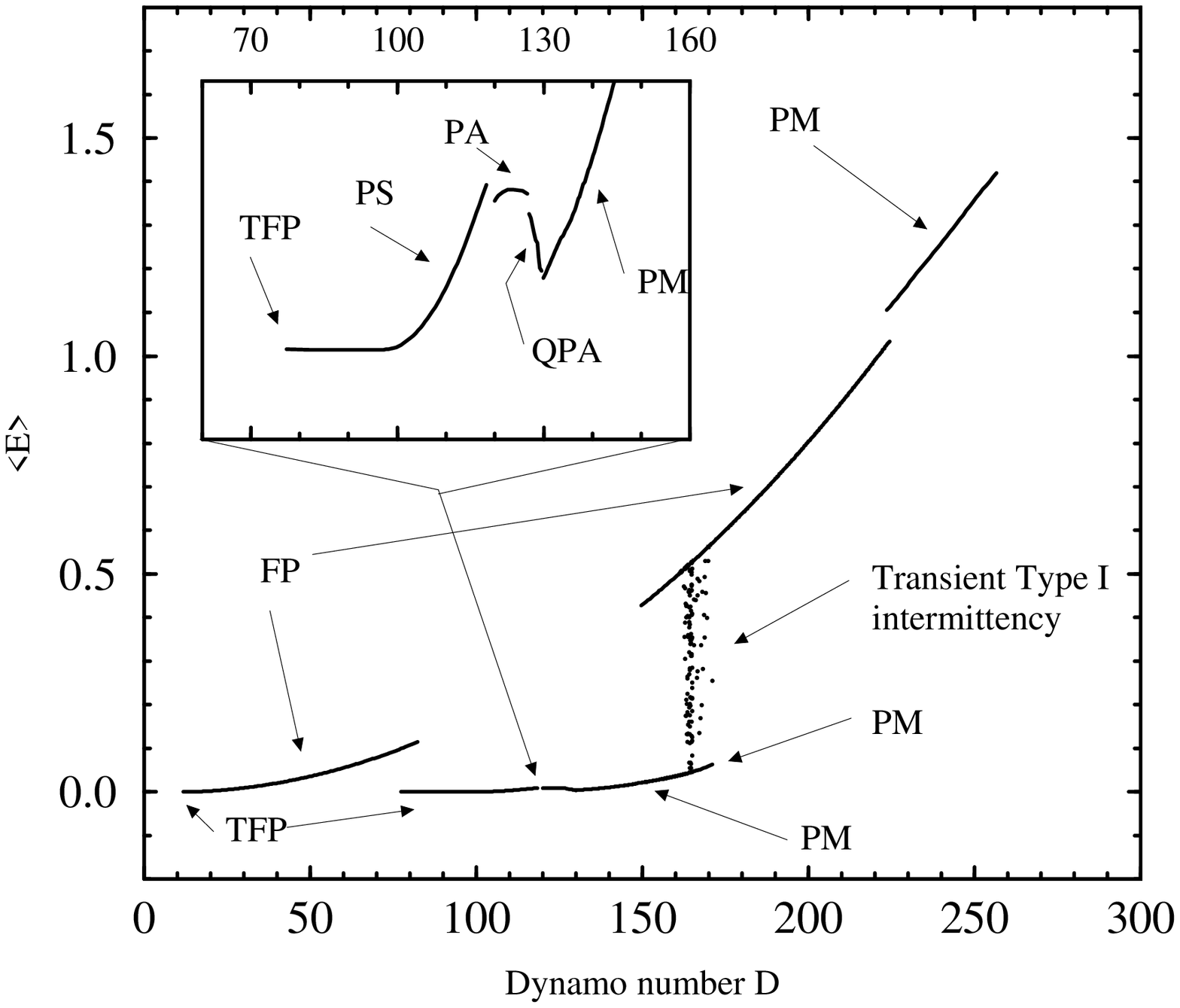}}
Figure 2

\centerline{\def\epsfsize#1#2{0.55#1}\epsffile{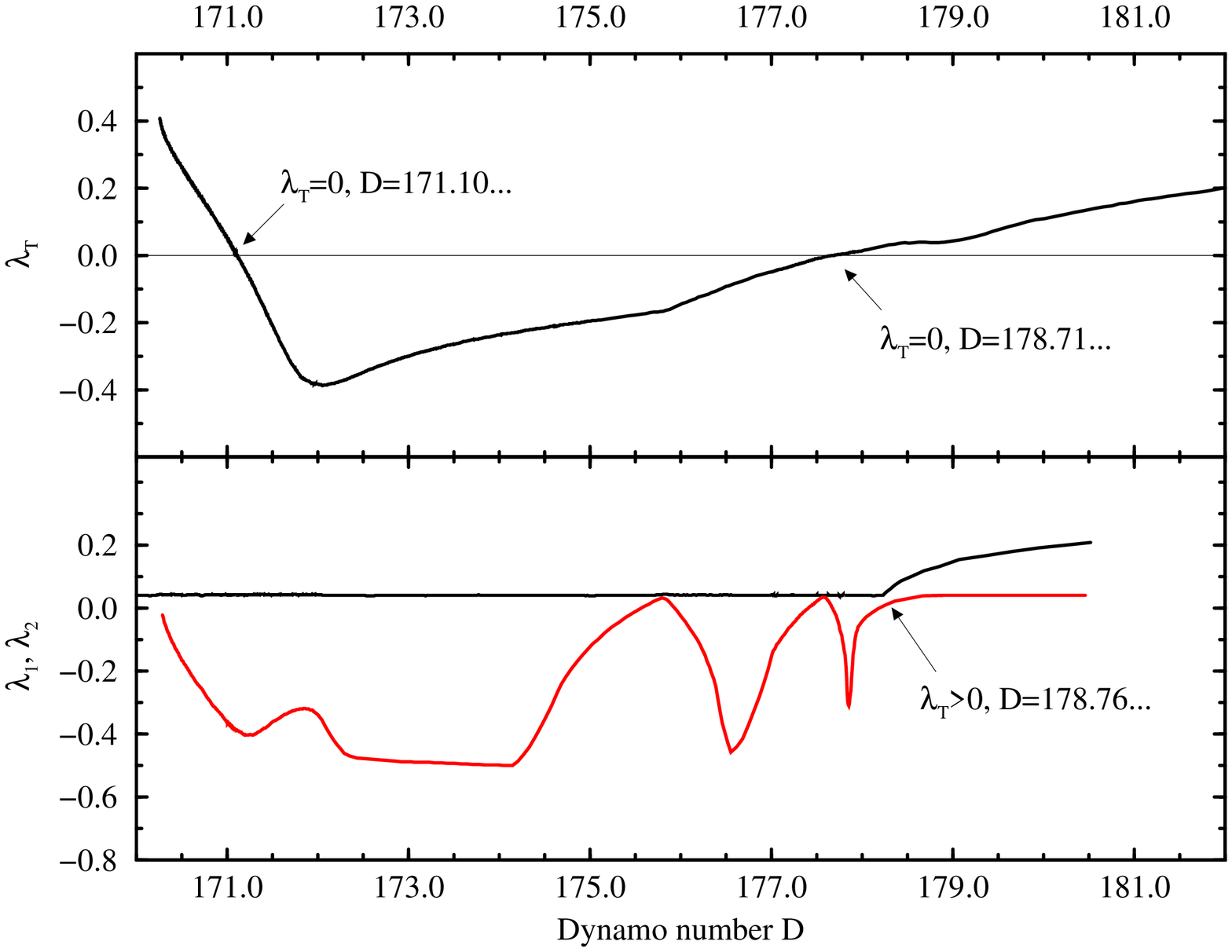}}
Figure 3

\centerline{\def\epsfsize#1#2{0.5#1}\epsffile{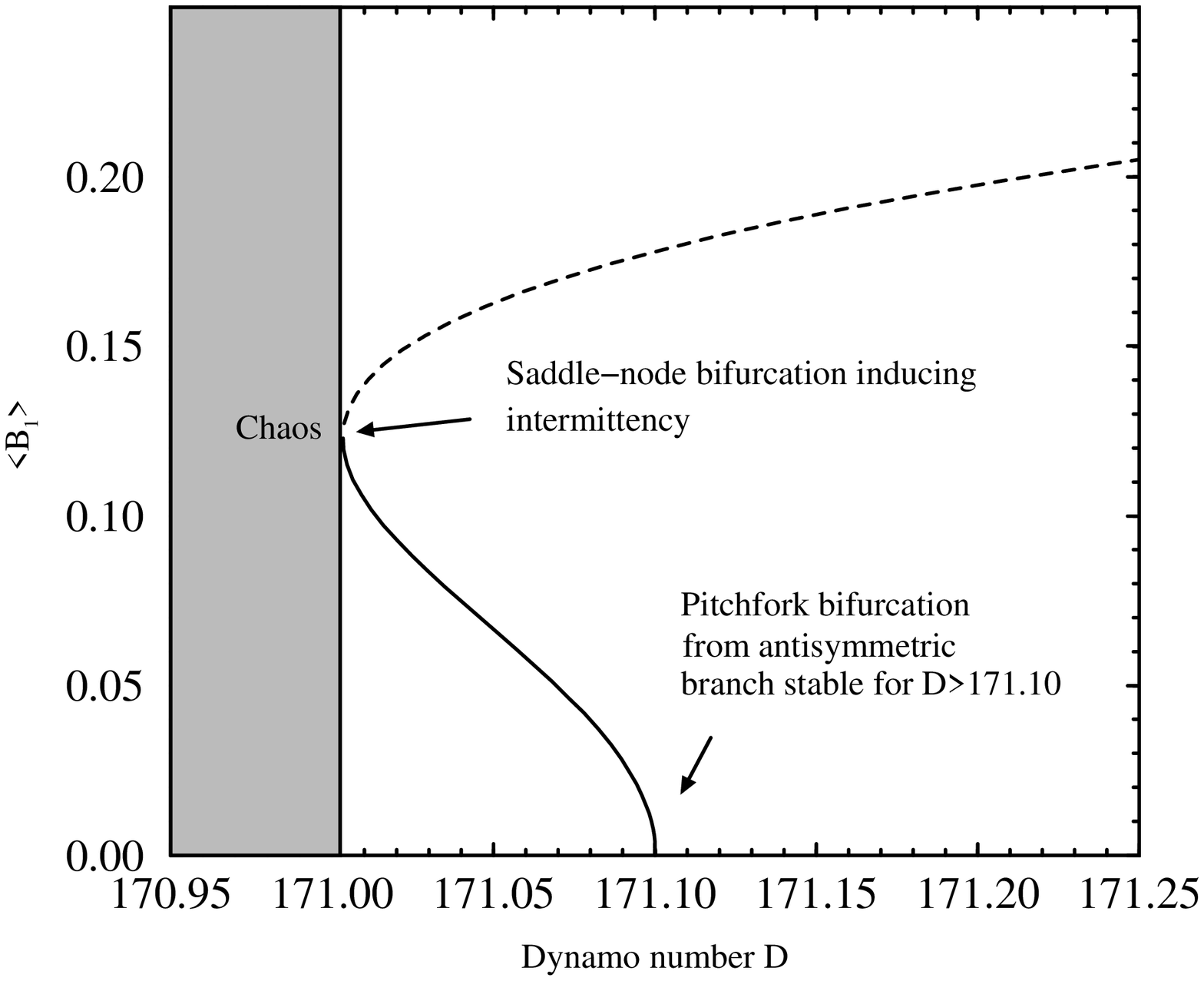}}
Figure 4

\centerline{\def\epsfsize#1#2{0.5#1}\epsffile{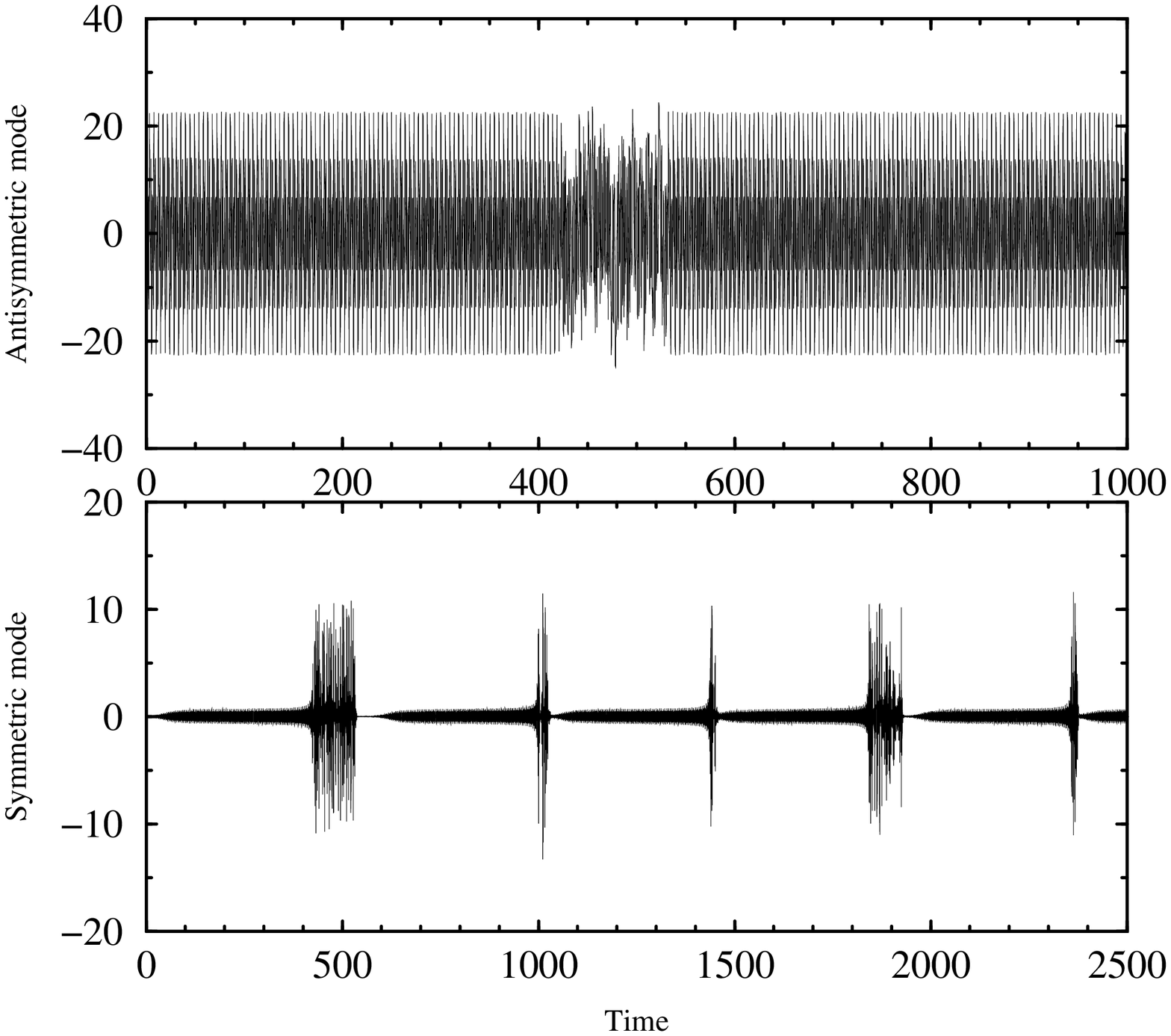}}
Figure 5

\centerline{\def\epsfsize#1#2{0.5#1}\epsffile{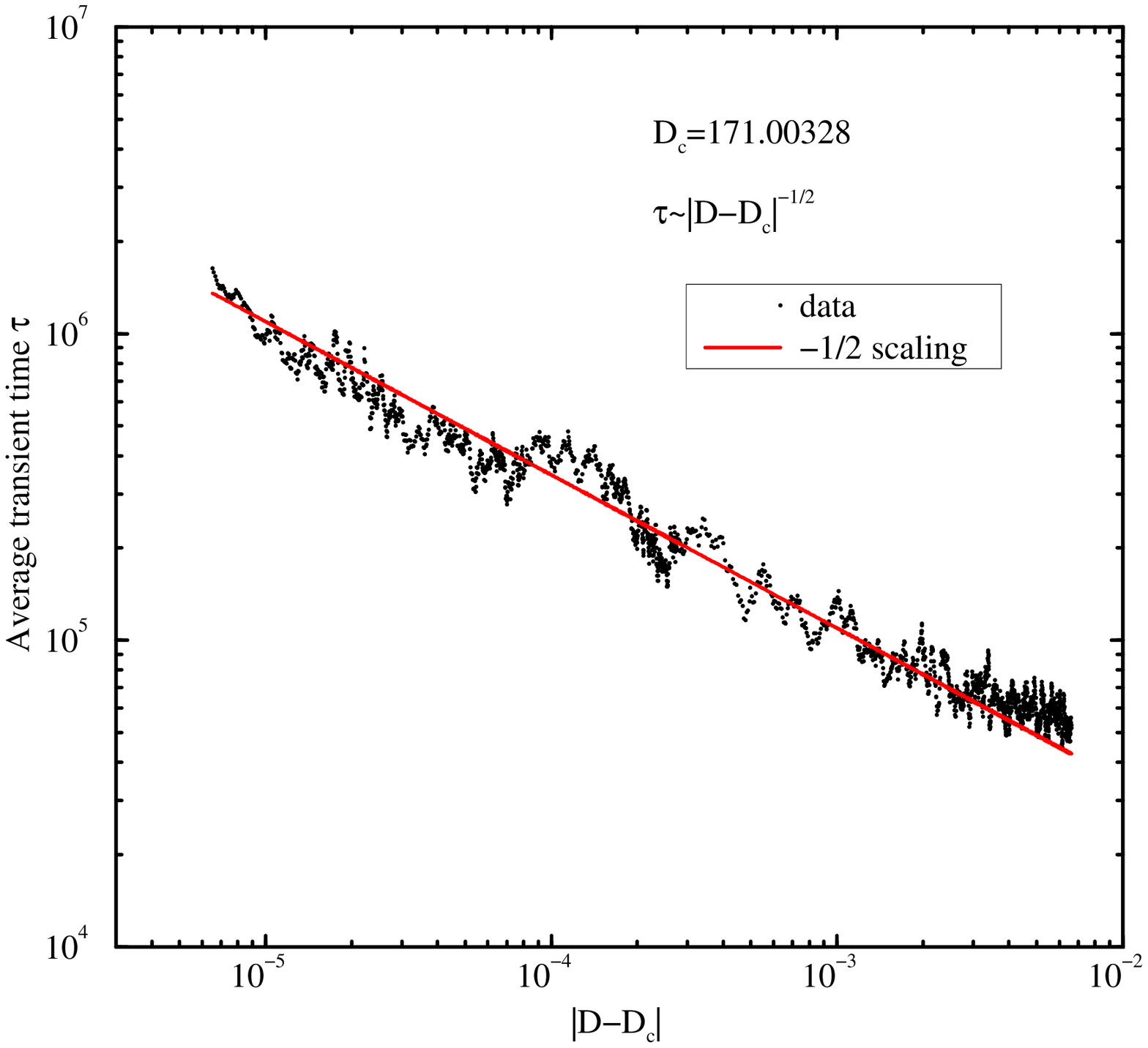}}
Figure 6

\centerline{\def\epsfsize#1#2{0.5#1}\epsffile{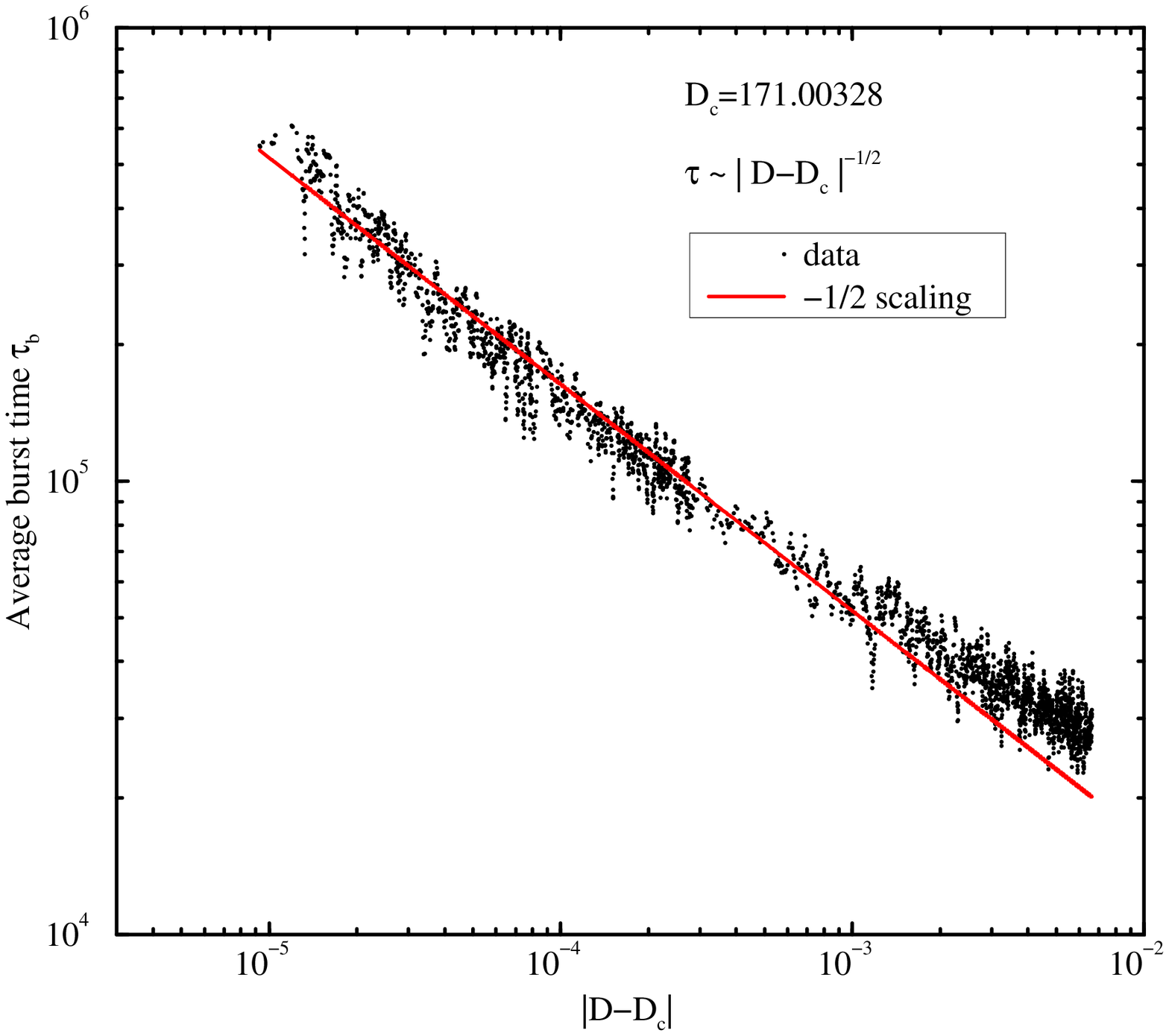}}
Figure 7

\centerline{\def\epsfsize#1#2{0.4#1}\epsffile{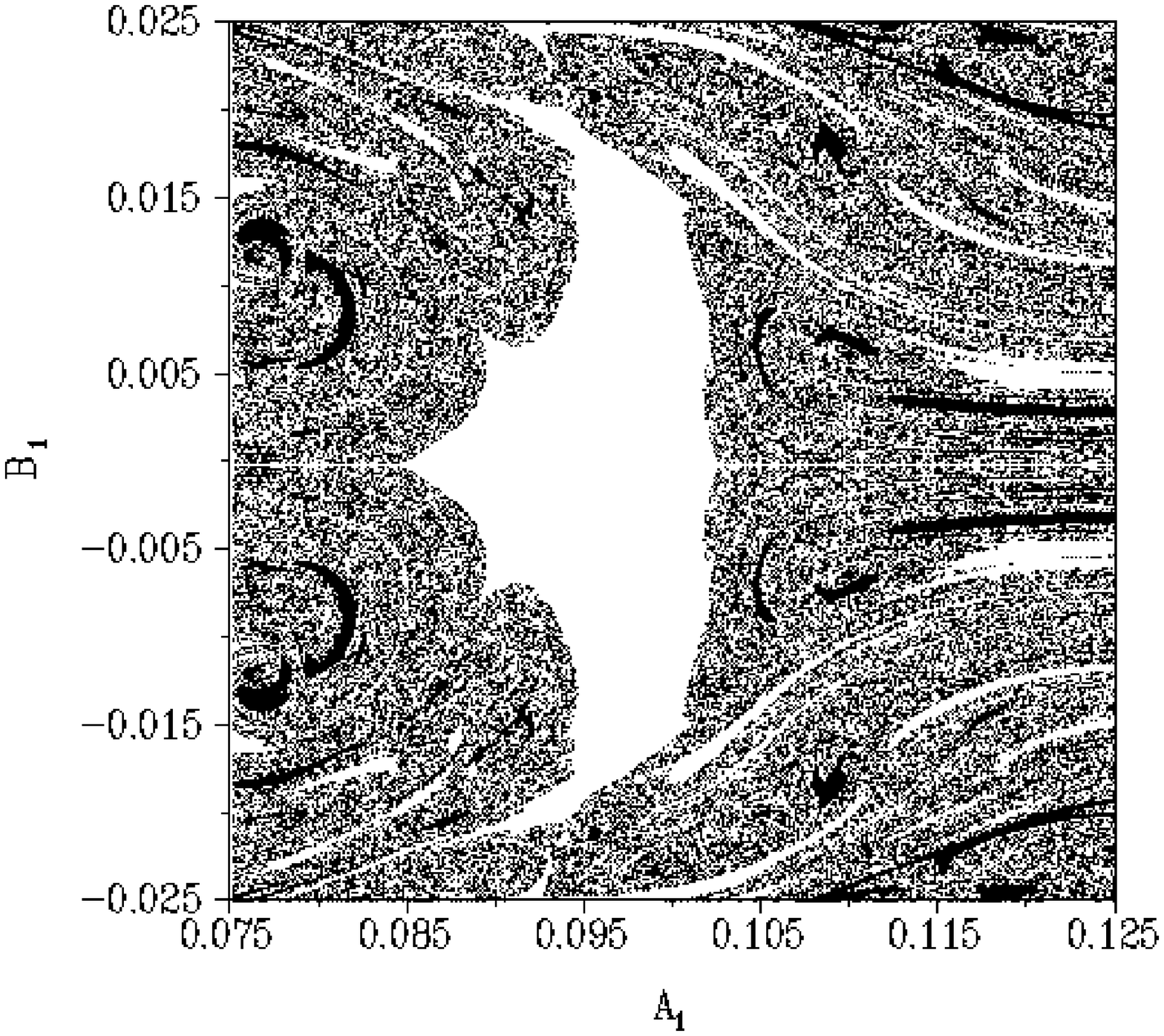}}
Figure 8

\centerline{\def\epsfsize#1#2{0.55#1}\epsffile{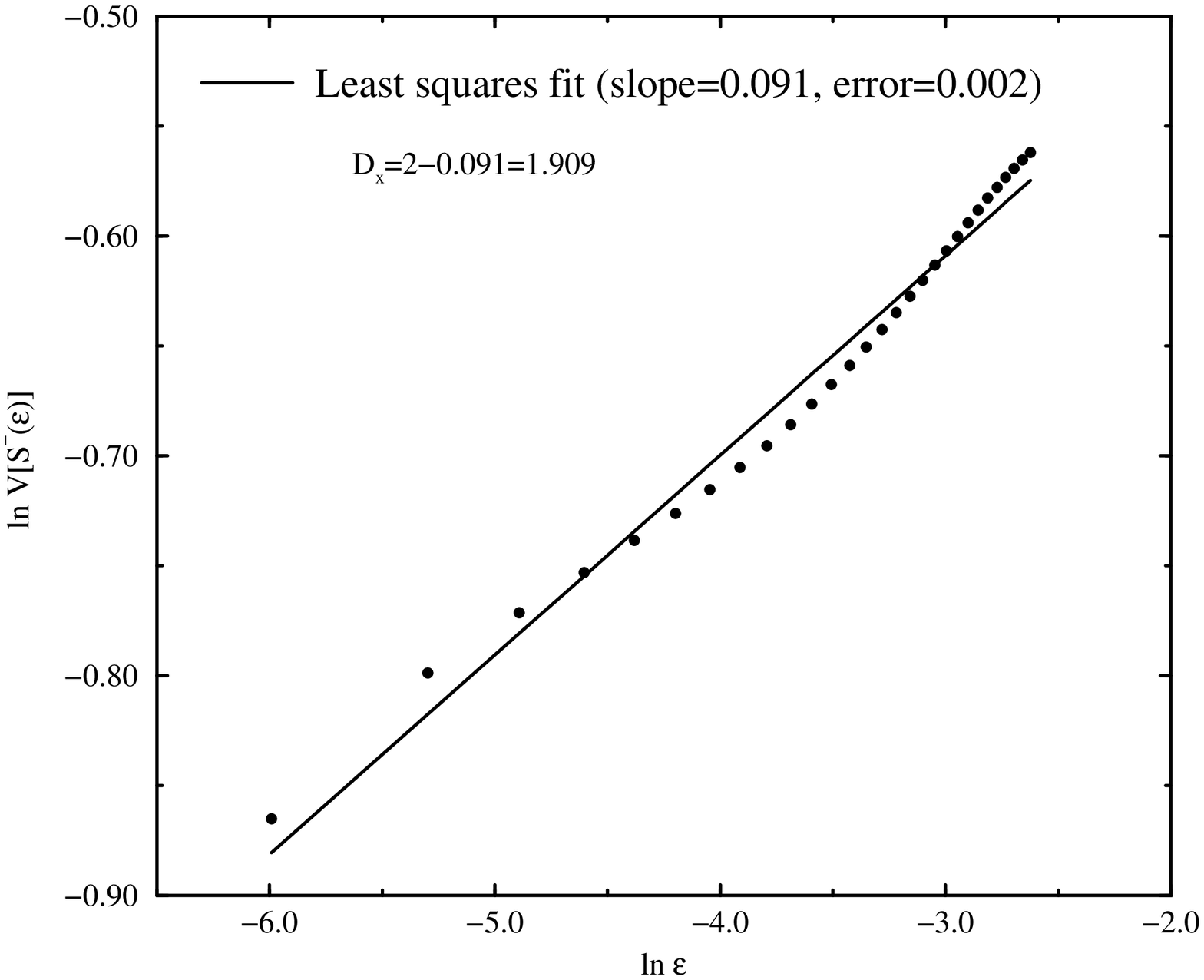}}
Figure 9

\centerline{\def\epsfsize#1#2{0.55#1}\epsffile{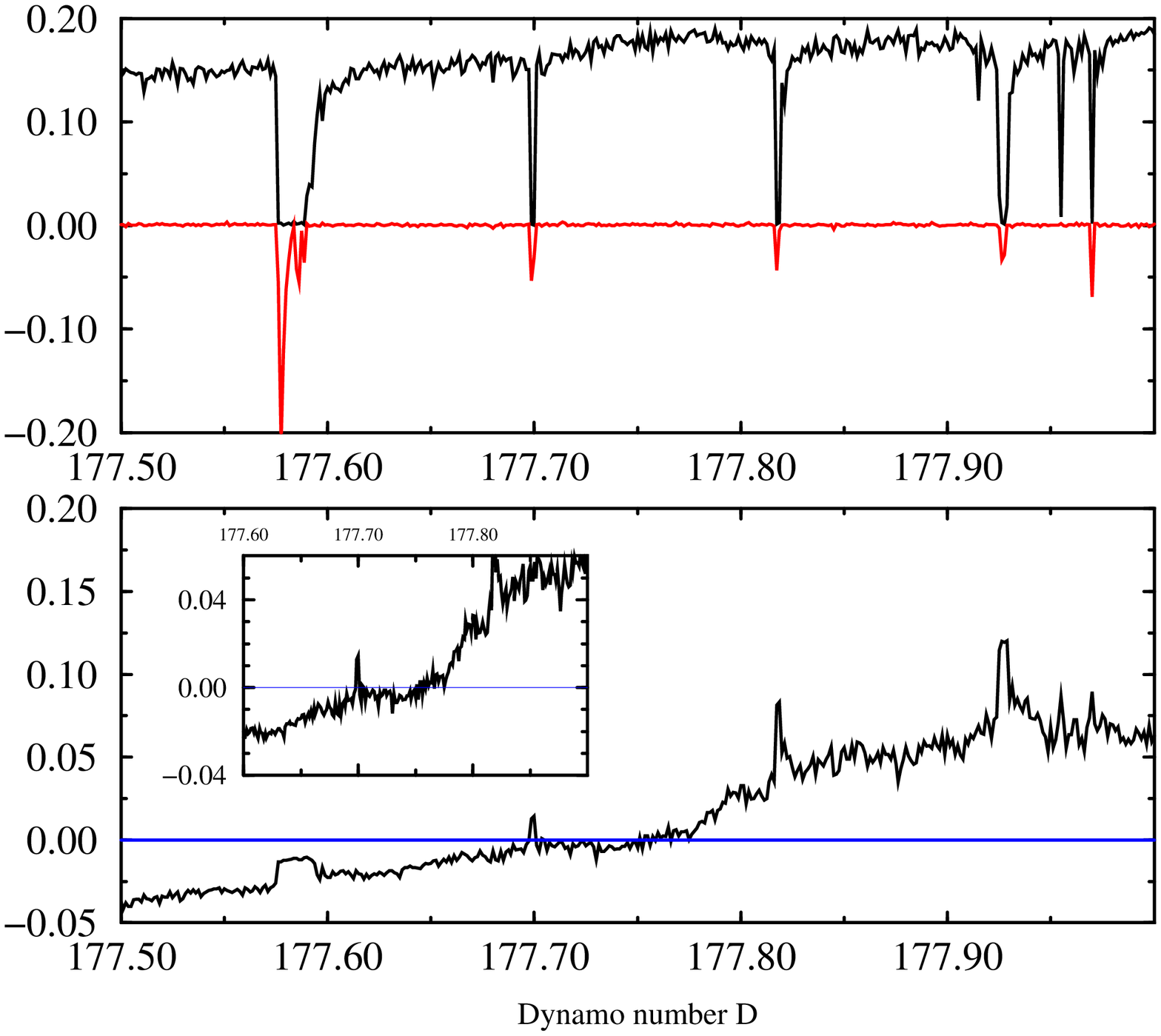}}
Figure 10

\centerline{\def\epsfsize#1#2{0.55#1}\epsffile{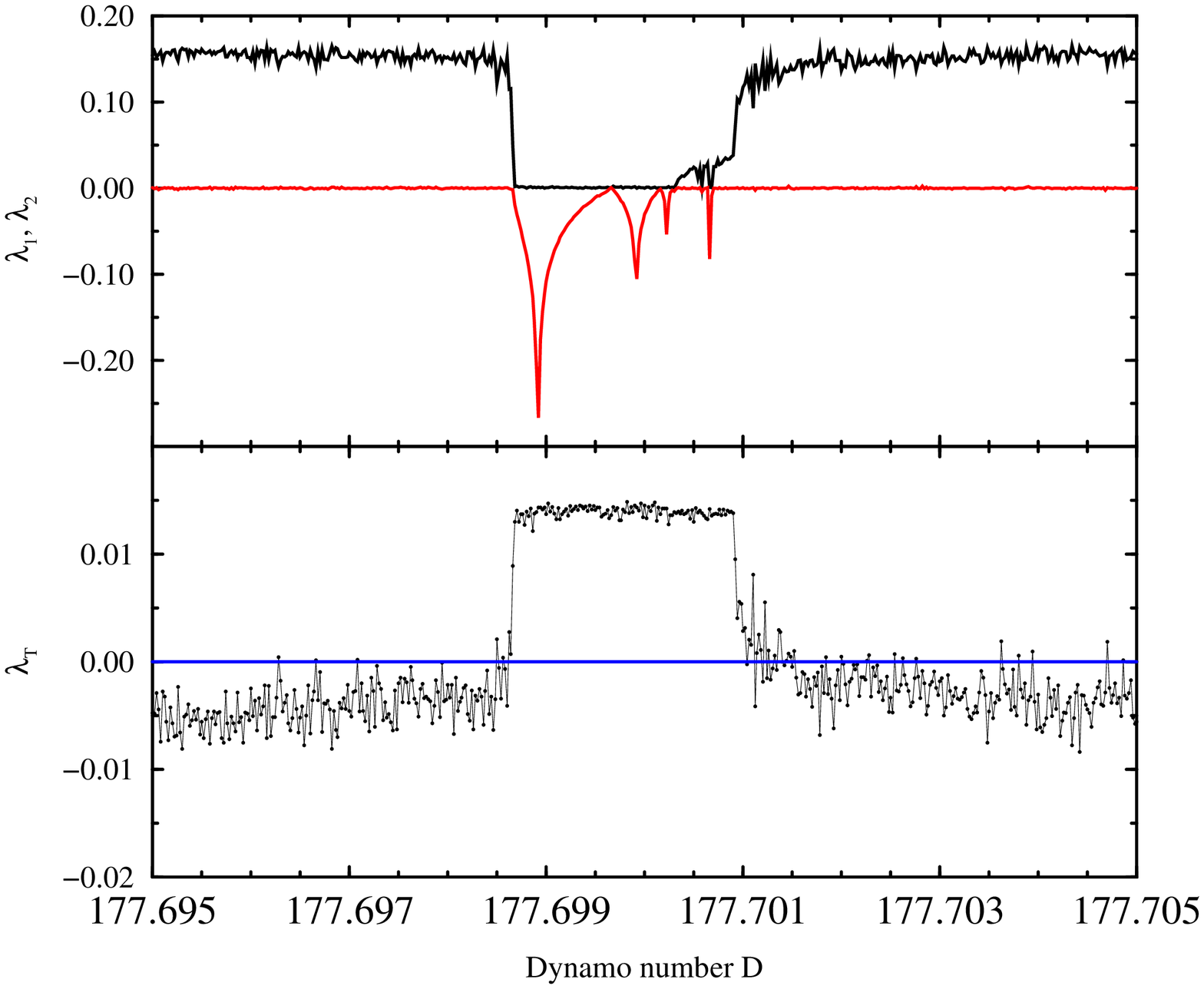}}
Figure 11

\centerline{\def\epsfsize#1#2{0.5#1}\epsffile{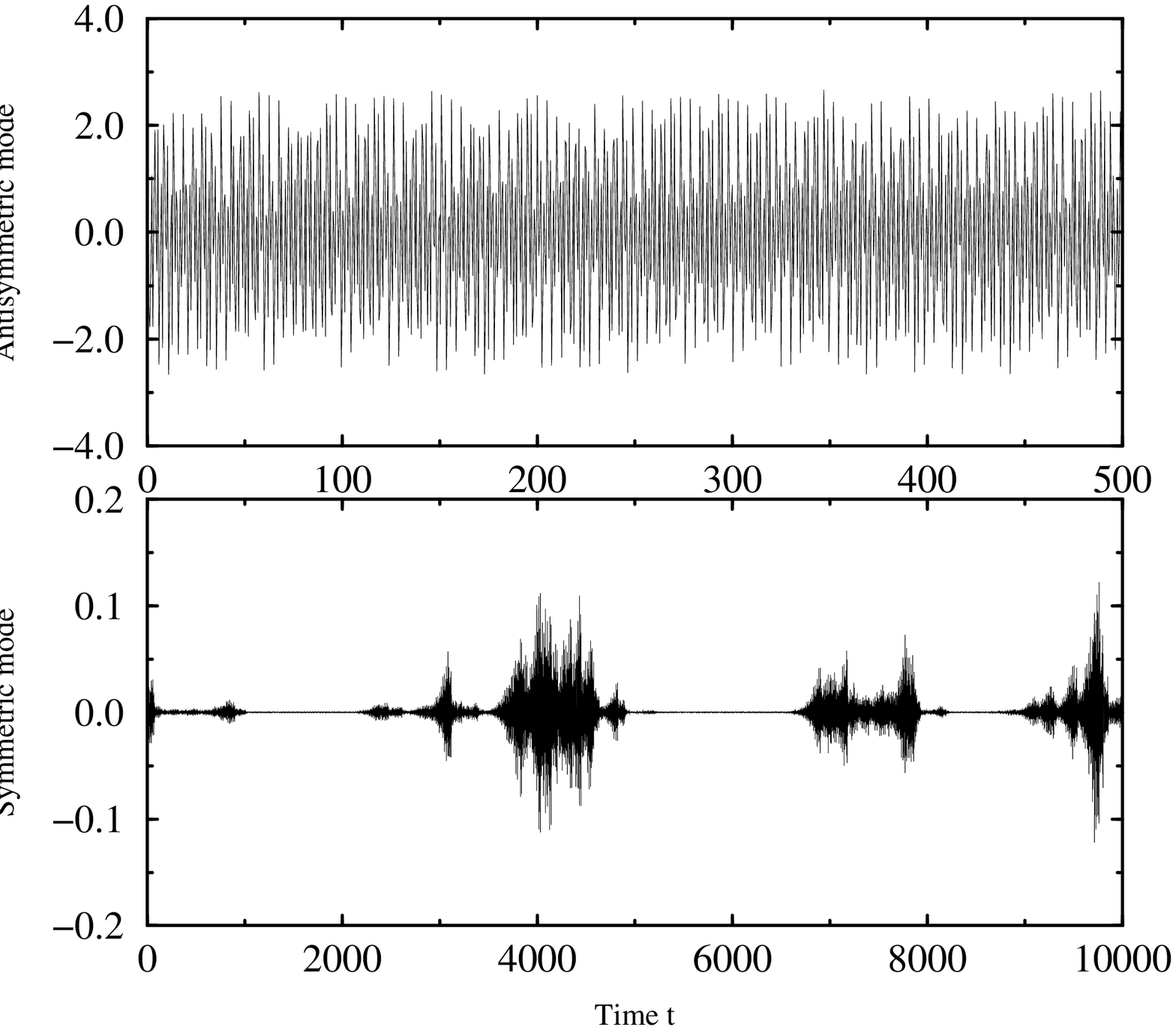}}
Figure 12

\centerline{\def\epsfsize#1#2{0.55#1}\epsffile{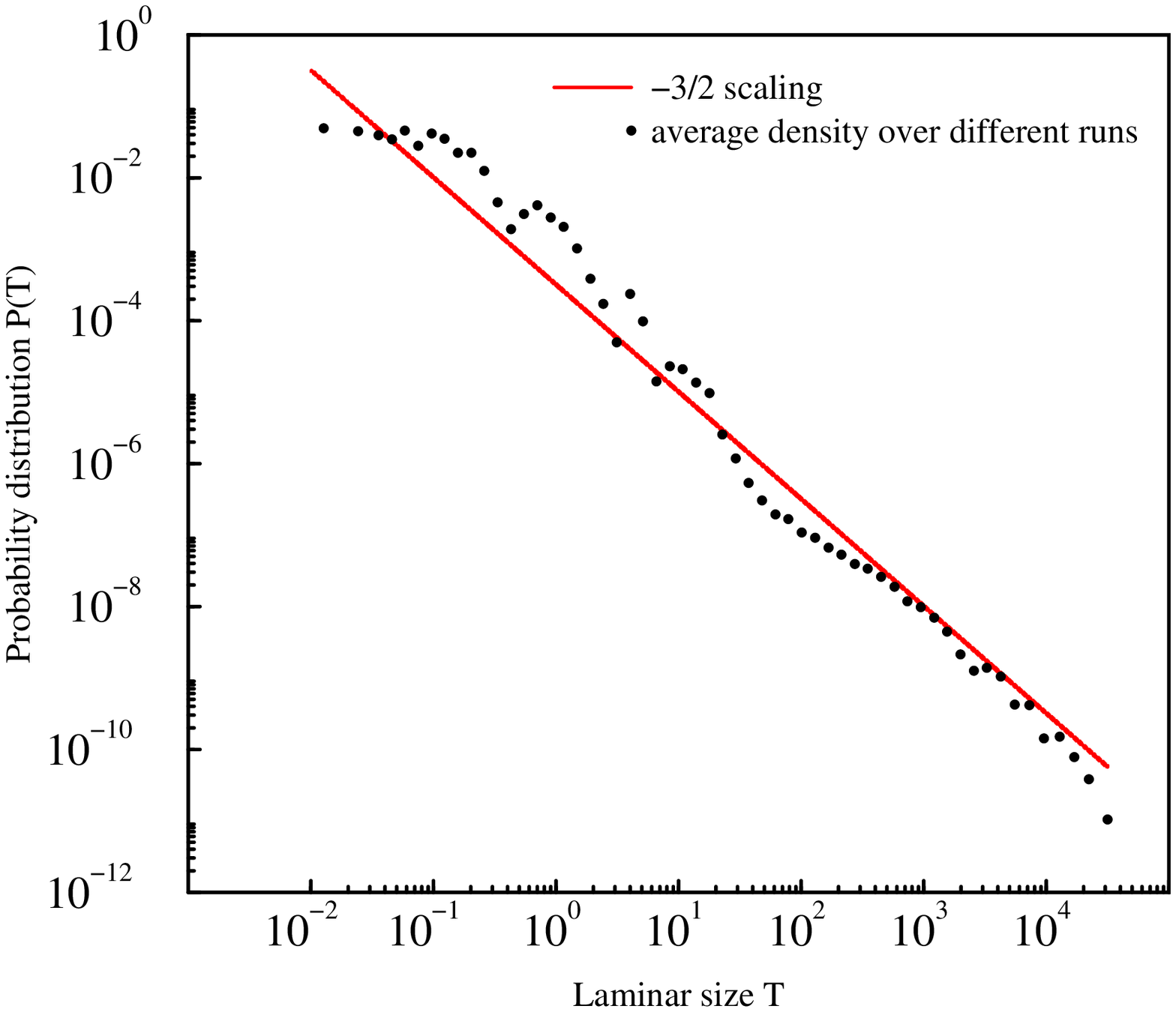}}
Figure 13

\centerline{\def\epsfsize#1#2{0.55#1}\epsffile{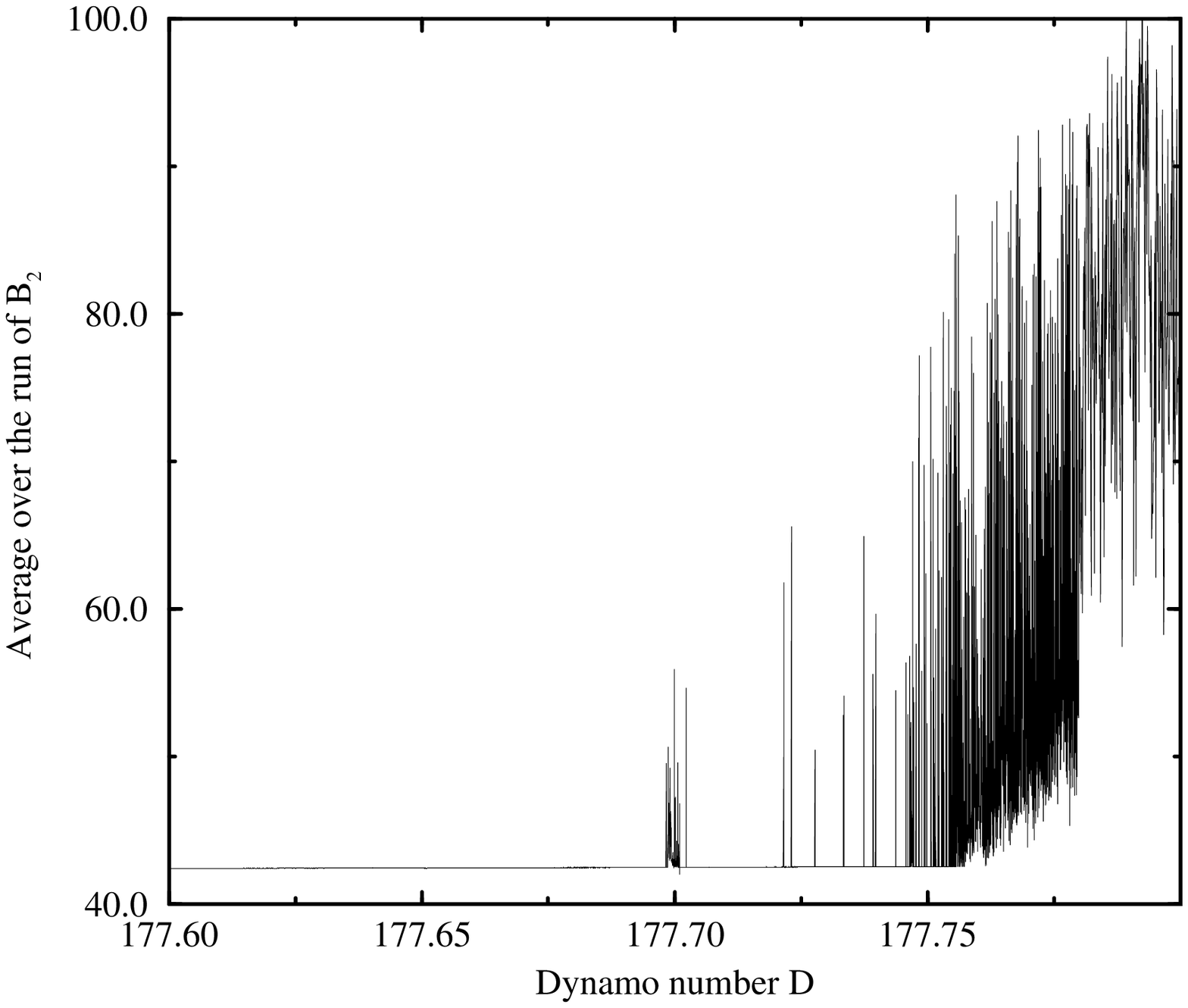}}
Figure 14

\end{document}